\documentclass[pre,aps,twocolumn,showpacs]{revtex4}
\usepackage{graphicx}
\usepackage{bm}
\newcommand{\B}[1]{{\bm{#1}}}

\usepackage[latin1]{inputenc}
\newcommand{\beq}{\begin{equation}}
\newcommand{\eeq}{\end{equation}}
\newcommand{\bea}{\begin{eqnarray}}
\newcommand{\eea}{\end{eqnarray}}

\begin{document}
\title{Drag Reduction by Microbubbles in Turbulent Flows: the Limit of Minute Bubbles}
\author{Victor S. L'vov, Anna Pomyalov, Itamar Procaccia and Vasil Tiberkevich}
\affiliation{ Dept. of Chemical
Physics, The Weizmann Institute of Science, Rehovot, 76100 Israel}

\begin{abstract}
Drag reduction by microbubbles is a promising engineering method
for improving ship performance. A fundamental theory of the
phenomenon is lacking however, making actual design quite
up-hazard. We offer here a theory of drag reduction by
microbubbles in the limit of very small bubbles, when the effect
of the bubbles is mainly to normalize the density and the
viscosity of the carrier fluid. The theory culminates with a
prediction of the degree of drag reduction given the concentration
profile of the bubbles. Comparisons with experiments are discussed
and the road ahead is sketched.

\end{abstract}
\maketitle

The idea of reducing drag friction by placing a thin layer of air
between a ship and its water boundary was patented already in the
nineteenth century \cite{patent}. Drag reduction by the injection
of microbubbles into the turbulent boundary layer  has been the
subject of intensive research since the first experimental
observation of this phenomenon in \cite{73BM}, and see the
comprehensive review \cite{90MD}. The reduction of skin-friction
drag by microbubbles has important technological and engineering
advantages, especially for marine transportation by huge and
relatively slow ships like tankers, but also for many other
applications, such as hydro-foils, in-pipe transportation, etc.
The voluminous literature on the engineering aspects of the
problem cannot be referenced in full. It suffices to mention
impressive results such as the microbubble drag reduction by about
80\% on a flat plane,~\cite{85MDM}, and up to 32\% on a 50 m long
flat plane ship, see~\cite{99TKMYK}. Some steps in understanding
the phenomenon  have been made. Ref~\cite{77BEM} found that the
drag reduction correlates with the maximum void fraction in the
boundary layer. It was understood that the ``local distribution
and shape [of the microbubble void fraction $C(\bm{r})$] in the
boundary layer have paramount influence in the drag
reduction"~\cite{02HV}. Many researches~(see, e.g.~\cite{98Wat})
found  that effect of micro-bubbles decreases downstream and that
the  bubble size is another important factor influencing
frictional resistance.

Legner~\cite{84Leg} stated that the ``decrease of the medium
density as the gas concentration increases provides the primary
drag reduction mechanism".  Unfortunately, the analysis of
Ref.~~\cite{84Leg} does not contain any spatial dependencies,
taking the distribution of bubble void fraction to be homogeneous.
In addition Legner~\cite{84Leg} concluded that the increase of the
dynamic fluid viscosity, caused by the bubbles, leads to
\emph{increase of frictional drag}. In contradiction, other
studies (see, e.g. Ref.~\cite{99Kat}) lead to the opposite
conclusion:  that the\emph{ increase} of the  viscosity, caused by
microbubbles, \emph{decreases} the friction drag. To date this
confusion has not been resolved theoretically.

The aim of this Letter is to offer a theory of drag reduction by
microbubbles in the limit that their diameter $d$ is very small
($d\to 0$), and the void fraction $C(\B r)$ is fixed, and not too
large ($C(\B r)\le 0.1$). In addition we will assume that the
scale of variation $\ell_C\equiv C(\B r)/|\B \nabla C(\B r)|\ll z$
where $z$ is the distance from the wall. The advantage of a theory
in this limit is that we can show quite rigorously that the only
mechanism for drag reduction available in this limit is provided
by the reduction of the fluid density and the increase in the
fluid viscosity. This is not to say that there are no additional
possible mechanisms of drag reduction by microbubbles due to their
influence on the structure of turbulence, including near wall
coherent structures ~\cite{02MK,04FE,04SKTM}. The theoretical
description of such effects is however very difficult; they stem
entirely from finite bubble-size effects, and they should be taken
only as a further step in the development of the theory.

As a starting point for the theoretical development we take the
two-fluid description of turbulent flows with bubbles which is
presented in Ref.~\cite{97ZP}. In this description the bubbles are
of diameter $d$ which is very small. We do not consider individual
bubbles, but rather describe them by a field of void fraction
$C(\B r,t)\ll 1$ and velocity $\bm{w}(\B r,t)$. The carrier fluid
has density $\rho_0$, viscosity $\mu_0$ and velocity $\bm{U}(\B
r,t)$. We will take the air density of the bubbles to be zero and
the acceleration due to gravity, $\bm{g}$,  to act in the $\hat z$
direction which is normal to the wall. Disregarding terms of the
order of $d^2$ one write the equation of motion
\begin{eqnarray}
&&(1-C)\rho_0  \frac{D \bm{U}}{D t} = \frac{18 \, C\, \mu_0}{
d^2}\left(\bm w -\bm U\right)-(1-C)\bm\nabla p
\label{newNS}\\&&+(1-C)\rho_0\bm g   +2(1-C)\bm\nabla\cdot\left(\mu \bm E\sb m\right)
-\frac{3}{4} \mu_0 C \nabla^2\bm U  \ , \nonumber \\
&&\hskip -0.5 cm 2\mu_0\bm\nabla\cdot\bm E\sb f -\bm\nabla p
- \frac{18\mu_0}{d^2}\left(\bm w -\bm U\right)
+\frac{3}{4}\mu_0\nabla^2\bm U =0 \ . \label{bub}
\end{eqnarray}
In these equations
\begin{eqnarray}
&& \frac{D}{D t}= \frac{\partial }{ \partial t}+
 \bm U\cdot\bm\nabla \ , \quad  \bm E\sb m \equiv \case{1}{2}\left[
\left(\bm\nabla\bm U\sb m\right)+\left(\bm\nabla\bm U\sb m\right)^T \right]\ ,\nonumber\\
&&\hskip -0.3 cm\bm U\sb m\equiv (1-C)\bm U +C\bm w\ ,
 \quad \bm E\sb f \equiv \case{1}{2}\left[
\left(\bm\nabla\bm U\right) +\left(\bm\nabla\bm U\right)^T \right]\ .
\nonumber
\end{eqnarray}
The effective viscosity which appears in these equations is
determined by the bubble concentration,
\begin{equation}
\label{mu}
\mu \equiv  \left(1+5\,C\,\big /2 \right)\mu_0\ .
\end{equation}
These equation should be supplemented with the continuity equations
\begin{eqnarray}\label{cont-u}
\partial(1-C)\big / \partial t
+\bm\nabla\cdot\left[(1-C)\bm U\right] &=& 0 \,,\\ \label{cont-w}
\partial C\big /\partial t +\bm\nabla\cdot\left(C\bm w\right)
&=& 0 \ .\end{eqnarray} We now simplify the equations further in
the limit $d\to0$ by evaluating the term proportional to $d^{-2}$
in Eq. (\ref{newNS}) using the same term in Eq. (\ref{bub}). We
find
\begin{eqnarray}
(1-C)\rho_0 \frac{D\bm U}{D t}&=&(1-C)\left[ -\bm\nabla p
+\rho_0\bm g +2\bm\nabla\cdot\left(\mu \bm E\sb m\right)\right]
\nonumber\\&& +C\left( -\bm\nabla p +2\mu_0\bm\nabla\cdot\bm E\sb f
\right) \ .
\end{eqnarray}
In the same limit $\bm U\sb m=\bm U$ and $\bm E\sb m=\bm E\sb f\equiv \bm{E}$.

After some further simplifications in which we retain only terms
linear in $C(\B r)$ one gets:
\begin{eqnarray}\label{one-fluid-a}
\rho\,\frac{D\bm U}{Dt} &=& -\bm\nabla p+\rho\bm g
+2\bm\nabla\cdot\left(\mu \bm E \right) \,,\\ \label{one-fluid-b}
\bm\nabla\cdot\bm U &=& 0\ , \quad  DC\big /Dt =  0 \ .
\label{one-fluid-c}
\end{eqnarray}
where the effective density of the suspension is
\begin{equation}\label{rho}
\rho \equiv (1-C)\rho_0
\ .\end{equation}
The important conclusion is that dilute ($C\ll 1$) solutions of
minute microbubbles ($d\to0$)  can be described by a one-fluid
model with modified density $\rho$ and viscosity $\mu$. Note that
velocity field remains incompressible; this result is valid for
minute microbubbles $d\to0$ for arbitrary concentrations $C$.
Having these results at hand we are poised to offer a theory of
drag reduction that is quite similar to the theory by the same
authors for drag reduction by flexible polymers \cite{04LPPT}.

Consider a flow in channel geometry (with half channel width $L$);
the mean flow is in the $x$ direction, the wall normal direction
is $z$ and the span-wise direction is $y$. We take the bubble
concentration $C(\B r)$ to be given and time independent. The
fluid velocity $\B U(\B r)$ is a sum of its average (over time)
and a fluctuating part:
\begin{equation}
\B U(\B r,t) = \B V(z) + \B u(\B r,t) \ , \quad \B V(z)
\equiv \langle \B U(\B r,t) \rangle \ .
\end{equation}
For channel flows all the averages, and in particular $\B V(z)
\Rightarrow V(z)$, are functions of $z$ only. The objects that
enter the theory are the mean shear $S(z)$, the Reynolds stress
$W(z)$ and the kinetic energy $K(z)$; these are defined
respectively as 
$$
 \!\! S(z)\! \equiv \! \frac{d V(z)}{d z}, \  W (z)\! \equiv\!  -
\rho(z)\langle u_xu_z\rangle, \   K(z) \! = \! \frac{ \rho(z)}2
\langle |\B u|^2\rangle \  .
$$
Under the assumption $\ell_c\ll y$ we derive point-wise balance
equation for the flux of mechanical momentum, relating  these
objects \cite{04LPPT,04LPT}. Near the wall  it reads:
\begin{equation}
 \mu (z)S(z) + W(z) = p'L \,, \quad \mbox{for } z\ll L\ .\label{MF}
\end{equation}
On the RHS of this equation we see the production of momentum
due to the pressure gradient; on the LHS we have the Reynolds
stress and the viscous contribution to the momentum flux, with the
latter being usually negligible (in Newtonian turbulence
$\mu=\mu_0$) everywhere except in the viscous boundary layer.

A second relation between $S(z)$, $W(z)$ and $K(z)$ is obtained
from the energy balance. The energy is created by the large scale
motions at a rate of $W(z) S(z)$. It is cascaded down the scales
by a flux of energy, and is finally dissipated at a rate
$\epsilon$, where $\epsilon = \mu(z) \langle |\nabla u|^2\rangle$.
We cannot calculate $\epsilon$ exactly, but we can estimate it
rather well at a point $z$ away from the wall. When viscous
effects are dominant, this term is estimated as $[\mu(z)/\rho(z)]
(a/z)^2 K(z)$ (the velocity is then rather smooth, the gradient
exists and can be estimated by the typical velocity at $z$ over
the distance from the wall). Here $a$ is a constant of the order
of unity. When the Reynolds number is large, the viscous
dissipation is the same as the turbulent energy flux down the
scales, which can be estimated as $K(z)/\tau(z)$ where $\tau(z)$
is the typical eddy turn over time at $z$. The latter is estimated
as $\sqrt{\rho(z)}z/b\sqrt{K(z)}$ where $b\sim 1$ is another
constant. We can thus write the energy balance equation at point
$z$ as
\begin{equation} \label{EB}
\left[\frac{\mu(z)}{\rho(z)}\left(\frac{a}{z}\right)^2
+\frac{b\sqrt{K(z)}}{\sqrt{\rho(z)}z}\right]K(z) = W(z)S(z)\,,
\end{equation}
where the bigger of the two terms on the LHS should win. We note
that contrary to Eq. (\ref{MF}) which is exact, Eq.(\ref{EB}) is
not exact. It was shown however to give excellent order of
magnitude estimates as far as drag reduction is concerned
\cite{04LPPT,04BLPT}.  Finally, we quote the experimental fact
\cite{pope,01PNVH} that outside the viscous boundary layer
\begin{equation}
\label{WK} W(z)=c^2 K(z) \ ,
\end{equation}
with the coefficient $c$ rigorously bounded from above by unity
(The proof is $|c|^2\equiv |W|/K\le 2|\langle u_xu_z\rangle|/
\langle u_x^2+u_z^2\rangle\le 1$, because $(u_x\pm u_z)^2\ge 0$).

We can change variables now in favor of wall units according to
\begin{eqnarray*}
S^+&\equiv& \mu(z)\, S \big/ {p'L}\,,  \quad z^+\equiv z
\sqrt{\rho(z)p'L}\big/ \mu(z) \ ,\label{+}\\
W^+&\equiv&  W \big /p'L\,,  \qquad K^+\equiv K \big / p'L \ .
\end{eqnarray*}
In these units our balance equations read
\begin{eqnarray}
&&S^++W^+=1 \ , \quad K^+=c^2W^+\label{++} \ ,\\
&&\left[\left(\frac{a}{z^+}\right)^2+\frac{b}{z^+}\sqrt{K^+}
\right] K^+=W^+S^+ \ .
\end{eqnarray}
This set of equations is readily solved, giving
\begin{eqnarray}
&&S^+(z^+)=1 \ , \quad \text{for}~ z^+\le z^+_v\ , \label{viscous}\\
&&S^+(z^+)=\frac{2\kappa^2(z\sb{v}^+)^2 -1
+\sqrt{4\kappa^2\left[(z^+)^2 -(z\sb v^+)^2\right] +1}
}{\displaystyle 2\kappa^2(z^+)^2} \ ,\nonumber\\
&&\hskip 2.5cm \text{for}~ z^+\ge z^+_v \ . \label{Splus}
\end{eqnarray}
In these equations we defined $\kappa \equiv c^3/b$, $ z^+_v\equiv
a/c$. The mean velocity anywhere in the channel can be obtained by
integrating,
\begin{equation}\label{v-1}
V(z) = \int_0^zS(z')dz'=
\int_0^z \frac{p'L}{\mu(z')}S^+(z^+(z'))dz' \ .
\end{equation}
A measure of drag reduction is the relative increase in the mean
centerline velocity in the bubbly flow with respect to the neat
Newtonian fluid:
\begin{equation}\label{DV}
\Delta V \equiv V_{\rm bub}(L) -V_{\rm N}(L) \ .
\end{equation}
Clearly, $\Delta V>0$ corresponds to the drag {\em reduction},
while $\Delta V<0$ to the drag {\em enhancement}. We obtain an
expression for $\Delta V$ from Eq. (\ref{v-1}) by expanding to
linear order in $C(z)$ (where our equations are valid anyway):
\begin{equation}\label{DVp1}
\Delta V^+ \equiv \Delta V \sqrt{\rho_0/p'L}=
\int_0^\infty\chi(z^+)C(z^+)dz^+ \ .
\end{equation}
Here the response function $\chi$ consist of two parts, one due to
the density variation $\chi_\rho$ and the other due to the
viscosity variation $\chi_\mu$:
\begin{eqnarray}\label{sl-a}
\chi(z^+) &=& \chi_\rho(z^+) +\chi_\mu(z^+) \,,\\ \label{sl-c}
\chi_\rho(z^+) &=& -\frac{z^+}{2} \frac{\partial S^+(z^+)}{\partial z^+}
\,,\\\label{sl-d}
\chi_\mu(z^+) &=&-\frac{5}{2} \frac{\partial \big[S^+(z^+)z^+\big]}
{ \partial z^+} \ .
\end{eqnarray}
In writing Eq. (\ref{DVp1}) we have used the fact that in
experiments the bubbles tend to be localized in a finite region
near the wall, i.e. $C(z)\to 0$ sufficiently fast as $z\to\infty$,
and we extended the integration range to infinity.
\begin{figure}\vskip -0.7cm
\centerline{\includegraphics[width=.52\textwidth]{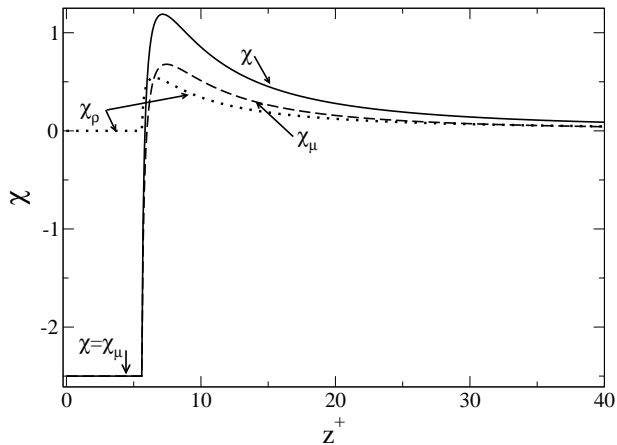}}
\vskip -0.4cm \caption{Plots of the model response function $\chi$
and its contributions $\chi_\rho$ and $\chi_\mu$ due to the
density and viscosity variations. These functions are computed
from our model shear function $S^+(z^+)$ as shown in Eqs.
(\ref{viscous}, \ref{Splus}) plugged into Eqs. (\ref{sl-a},
\ref{sl-c}). } \label{chi}
\end{figure}

Eqs. (\ref{DVp1})-(\ref{sl-d}) are the main theoretical
predictions of this Letter. To complement the theory we present
now estimates of the numerical value of the expected drag
reduction, and compare it with a relevant experiment. The simplest
model takes the parameters in Eq. (\ref{params}) as
$z$-independent, and in agreement with the classical von
K\'arm\'an boundary layer theory, i.e. $\kappa\approx 0.436$ and
$z_v^+\approx 5.6$. Evaluating the response function $\chi$ with
these parameters results with the findings presented in
Fig.~\ref{chi}. We see that in the viscous layer, where Eq.
(\ref{viscous}) is relevant, $\chi_\rho\approx 0$ while $\chi_\mu$
is negative. This means that having a bubble concentration in this
region does not buy us drag reduction due to the density
variation, but it leads to drag enhancement due to the viscosity
increase. This is far from being surprising, since in this region
the momentum flux is dominated by the viscous term $\mu S$. For a
fixed momentum flux any increase in viscosity must decrease $S$
and correspondingly lead to drag enhancement. The most efficient
drag reduction can be obtained by placing the bubble concentration
out of the viscous layer, but not too far from the wall, say at
$6\le z^+\le 30$. In this region both the decrease in density and
the increase in viscosity lead to drag reduction. The momentum in
this  region is transported mainly by the Reynolds stress
$-\rho\langle u_xu_z\rangle$. The effect of density reduction is
absolutely clear: it leads to the reduction in momentum flux. For
a given momentum generation $p'L$ this has to result in the
increase of the mean momentum of the flow.  More interesting and
counter-intuitive is the effect of increasing viscosity. In order
to understand it, we remind the reader that for intermediate
values of $z^+$ there is no well-developed turbulent cascade, and
outer and inner scales of turbulence are of the same order of
magnitude. Therefore the increase of viscosity reduces the
turbulent energy, in contrast to fully developed turbulence where
changes of viscosity simply modify the Kolmogorov scale without
any effect on the turbulent energy, that is dominated by the outer
scale. The decrease in turbulent energy here reduces the Reynolds
stress, see Eq. (\ref{WK}). It is interesting to note that this
effect of increasing viscosity is essentially the same as the
mechanism for drag reduction in the case of elastic
polymers~\cite{04LPPT,04BLPT}.  For polymers, however, the
increase in viscosity can be very significant and the linear
approximation that is used here is not  applicable.

In comparing with experiments we need to consider low bubble
concentrations. An interesting experiment was reported in
\cite{99Kat}, where both $C(z)$ and the $V^+(z^+)$ are shown. We
note that this experiment deals with a developing boundary layer
rather than a steady channel geometry, but near the wall the
Reynolds number can be considered rather time-independent.
Digitizing the published profiles $C(z)$ and integrating them
numerically against our function $\chi(z)$ we obtain results for
$\Delta V$ which appear in good agreement with the data of
\cite{99Kat}, as long as $C(z)$ is small, $C\le 0.1$. For the two
lowest values of bubble concentration we agree with the data to
within 10-20\%, which is definitely within the experimental error
bars. For higher values of $C(z)$ the results of the experiment
become sensitive to nonlinear effects.

\begin{figure}
\centerline{\includegraphics[width=.48\textwidth]{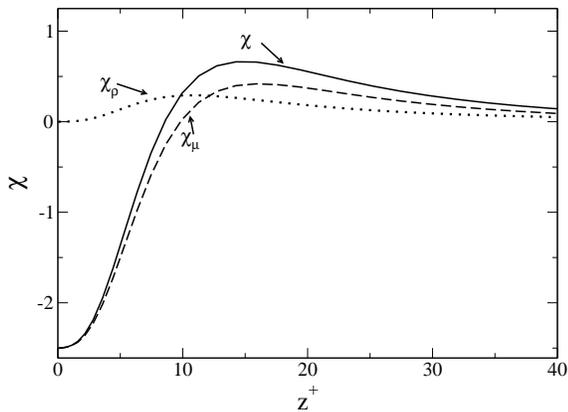}}
\caption{Plots of the response function $\chi$ and its
contributions $\chi_\rho$ and $\chi_\mu$ due to the density and
viscosity variations computed from the simulation data of
\cite{99MKM}). } \label{chinum}
\end{figure}

It appears extremely worthwhile to test the theory presented here
by numerical simulations that would be designed to do so. We
should stress that a careful measurement of $S^+(z^+)$ in either
experiments or simulations, in addition to a determination of
$\Delta V$, can provide a very good test of our theory. Eq.
(\ref{DVp1}) is more general than our model (\ref{Splus}), and it
can be tested directly if $S^+(z^+)$ and its $z^+$ derivative are
known. Since the response function $\chi$ is a property of the
reference (Newtonian) flow, we can take it from Newtonian data. As
an example of such a calculation we have considered the results of
numerical simulations for a Newtonian channel flow available in
\cite{99MKM}, where the profile $S^+(z^+)$ is provided. We have
used it to compute the response function $\chi$ and its two
contributions according to Eqs. (\ref{sl-a})-(\ref{sl-c}). The
results are presented in Fig. \ref{chinum}. We see that the
qualitative predictions of our model for $\chi$ are excellently
reproduced by the numerical data, even though the smoother cross
over between the viscous and logarithmic layers translates to
smoother functions $\chi_\rho$ and $\chi_\mu$. A similar
comparison for channel flow with bubbles will shed important
additional light on our approach.

We reiterate that additional nonlinear contributions to drag
reduction are expected to come in when the concentration
increases, and especially when the bubble diameter $d$ increases.
One should definitely examine theoretically the nonlinear and
finite size effects and incorporate them into a more complete
theory of drag reduction by microbubbles. It is the proposition of
this Letter however that the limit $C(z)\ll 1$ and $d\to 0$ is a
relevant limit where the theory simplifies considerably and where
experiments, and especially numerical simulations, can give
valuable support for the present theory. It is important to
exhaust the linear effects of drag reduction by minute
microbubbles before landing on the much more involved nonlinear
theory.

\acknowledgements We thank Roberto Benzi and  Said  Elghobashi for
useful exchanges. This work has been supported in part by the
US-Israel Bi-National Science Foundation, by the European
Commission through a TMR grant, and by the Minerva Foundation,
Munich, Germany.

\end{document}